
%
%
\font\tenbf=cmbx10

\font\eightrm=cmr8
\font\eightit=cmti8
\font\germ=eufm10
\def\sectiontitle#1\par{\vskip0pt plus.1\vsize\penalty-250
 \vskip0pt plus-.1\vsize\bigskip\vskip\parskip
 \message{#1}\leftline{\tenbf#1}\nobreak\vglue 5pt}
\hsize=5.0truein
\vsize=7.8truein
\parindent=15pt
\nopagenumbers
\baselineskip=10pt
\line{\eightrm
Preliminary version
\hfil}
\vglue 5pc
\baselineskip=13pt
\headline{\ifnum\pageno=1\hfil\else%
{\ifodd\pageno\rightheadline \else \leftheadline\fi}\fi}
\def\rightheadline{\hfil\eightit
Fusion algebra and Verlinde's formula
\quad\eightrm\folio}
\def\leftheadline{\eightrm\folio\quad
\eightit
Anatol N. Kirillov
\hfil}
\voffset=2\baselineskip
\centerline{\tenbf
FUSION \hskip 0.1cm ALGEBRA \hskip 0.1cm AND \hskip 0.1cm
VERLINDE'S \hskip 0.1cm FORMULA }
\vglue 24pt
\centerline{\eightrm
ANATOL N. KIRILLOV
}
\baselineskip=12pt
\centerline{\eightit
Isaac Newton Institute for Mathematical Sciences,
}
\baselineskip=10pt
\centerline{\eightit
20 Clarkson Road, Cambridge, CB3 OEH, U.K.
}
\baselineskip=12pt
\centerline{\eightit
and }
\baselineskip=12pt
\centerline{\eightit
Steklov Mathematical Institute,
}
\baselineskip=10pt
\centerline{\eightit
Fontanka 27, St.Petersburg, 191011, Russia
}
\vglue 20pt
\centerline{  }
\centerline{\eightrm ABSTRACT}
{\rightskip=1.5pc
\leftskip=1.5pc
\eightrm\parindent=1pc
We show that the coefficients of a decomposition into an irreducible
components of the tensor powers of level $r$ symmetric algebra of adjoint
representation coincide with the Verlinde numbers. Also we construct (for
$sl(2)$) the representations of a general linear group those dimensions are
given by corresponding Verlinde's numbers.
\vglue12pt}
\baselineskip=13pt
\overfullrule=0pt
\font\germ=eufm10
\def\gt g{\hbox{\germ g}}
\def\qed{\hfill$\vrule height 2.5mm width 2.5mm depth 0mm$}
\def\m@th{\mathsurround=0pt}

\def\fsquare(#1,#2){
\hbox{\vrule$\hskip-0.4pt\vcenter to #1{\normalbaselines\m@th
\hrule\vfil\hbox to #1{\hfill$\scriptstyle #2$\hfill}\vfil\hrule}$\hskip-0.4pt
\vrule}}

\def\addsquare(#1,#2){\hbox{$
	\dimen1=#1 \advance\dimen1 by -0.8pt
	\vcenter to #1{\hrule height0.4pt depth0.0pt\vss%
	\hbox to #1{\hss{%
	\vbox to \dimen1{\vss%
	\hbox to \dimen1{\hss$\scriptstyle~#2~$\hss}%
	\vss}\hss}%
	\vrule width0.4pt}\vss%
	\hrule height0.4pt depth0.0pt}$}}

\def\Fsquare(#1,#2){
\hbox{\vrule$\hskip-0.4pt\vcenter to #1{\normalbaselines\m@th
\hrule\vfil\hbox to #1{\hfill$#2$\hfill}\vfil\hrule}$\hskip-0.4pt
\vrule}}

\def\Addsquare(#1,#2){\hbox{$
	\dimen1=#1 \advance\dimen1 by -0.8pt
	\vcenter to #1{\hrule height0.4pt depth0.0pt\vss%
	\hbox to #1{\hss{%
	\vbox to \dimen1{\vss%
	\hbox to \dimen1{\hss$~#2~$\hss}%
	\vss}\hss}%
	\vrule width0.4pt}\vss%
	\hrule height0.4pt depth0.0pt}$}}

\def\hfourbox(#1,#2,#3,#4){%

\fsquare(0.3cm,#1)\addsquare(0.3cm,#2)\addsquare(0.3cm,#3)\addsquare(0.3cm,#4)}

\def\Hfourbox(#1,#2,#3,#4){%

\Fsquare(0.4cm,#1)\Addsquare(0.4cm,#2)\Addsquare(0.4cm,#3)\Addsquare(0.4cm,#4)}

\def\hthreebox(#1,#2,#3){%
	\fsquare(0.3cm,#1)\addsquare(0.3cm,#2)\addsquare(0.3cm,#3)}

\def\htwobox(#1,#2){%
	\fsquare(0.3cm,#1)\addsquare(0.3cm,#2)}

\def\vfourbox(#1,#2,#3,#4){%
	\normalbaselines\m@th\offinterlineskip
	\vtop{\hbox{\fsquare(0.3cm,#1)}
	      \vskip-0.4pt
	      \hbox{\fsquare(0.3cm,#2)}
	      \vskip-0.4pt
	      \hbox{\fsquare(0.3cm,#3)}
	      \vskip-0.4pt
	      \hbox{\fsquare(0.3cm,#4)}}}

\def\Vfourbox(#1,#2,#3,#4){%
	\normalbaselines\m@th\offinterlineskip
	\vtop{\hbox{\Fsquare(0.4cm,#1)}
	      \vskip-0.4pt
	      \hbox{\Fsquare(0.4cm,#2)}
	      \vskip-0.4pt
	      \hbox{\Fsquare(0.4cm,#3)}
	      \vskip-0.4pt
	      \hbox{\Fsquare(0.4cm,#4)}}}

\def\vthreebox(#1,#2,#3){%
	\normalbaselines\m@th\offinterlineskip
	\vtop{\hbox{\fsquare(0.3cm,#1)}
	      \vskip-0.4pt
	      \hbox{\fsquare(0.3cm,#2)}
	      \vskip-0.4pt
	      \hbox{\fsquare(0.3cm,#3)}}}

\def\vtwobox(#1,#2){%
	\normalbaselines\m@th\offinterlineskip
	\vtop{\hbox{\fsquare(0.3cm,#1)}
	      \vskip-0.4pt
	      \hbox{\fsquare(0.3cm,#2)}}}

\def\Hthreebox(#1,#2,#3){%
	\Fsquare(0.4cm,#1)\Addsquare(0.4cm,#2)\Addsquare(0.4cm,#3)}

\def\Htwobox(#1,#2){%
	\Fsquare(0.4cm,#1)\Addsquare(0.4cm,#2)}

\def\Vthreebox(#1,#2,#3){%
	\normalbaselines\m@th\offinterlineskip
	\vtop{\hbox{\Fsquare(0.4cm,#1)}
	      \vskip-0.4pt
	      \hbox{\Fsquare(0.4cm,#2)}
	      \vskip-0.4pt
	      \hbox{\Fsquare(0.4cm,#3)}}}

\def\Vtwobox(#1,#2){%
	\normalbaselines\m@th\offinterlineskip
	\vtop{\hbox{\Fsquare(0.4cm,#1)}
	      \vskip-0.4pt
	      \hbox{\Fsquare(0.4cm,#2)}}}

\def\twoone(#1,#2,#3){%
	\normalbaselines\m@th\offinterlineskip
	\vtop{\hbox{\htwobox({#1},{#2})}
	      \vskip-0.4pt
	      \hbox{\fsquare(0.3cm,#3)}}}

\def\Twoone(#1,#2,#3){%
	\hbox{
	\normalbaselines\m@th\offinterlineskip
	\vtop{\hbox{\Htwobox({#1},{#2})}
	      \vskip-0.4pt
	      \hbox{\Fsquare(0.4cm,#3)}}}}

\def\threeone(#1,#2,#3,#4){%
	\normalbaselines\m@th\offinterlineskip
	\vtop{\hbox{\hthreebox({#1},{#2},{#3})}
	      \vskip-0.4pt
	      \hbox{\fsquare(0.3cm,#4)}}}

\def\Threeone(#1,#2,#3,#4){%
	\normalbaselines\m@th\offinterlineskip
	\vtop{\hbox{\Hthreebox({#1},{#2},{#3})}
	      \vskip-0.4pt
	      \hbox{\Fsquare(0.4cm,#4)}}}

\def\Threetwo(#1,#2,#3,#4,#5){%
	\normalbaselines\m@th\offinterlineskip
	\vtop{\hbox{\Hthreebox({#1},{#2},{#3})}
	      \vskip-0.4pt
	      \hbox{\Htwobox({#4},{#5})}}}

\def\threetwo(#1,#2,#3,#4,#5){%
	\normalbaselines\m@th\offinterlineskip
	\vtop{\hbox{\hthreebox({#1},{#2},{#3})}
	      \vskip-0.4pt
	      \hbox{\htwobox({#4},{#5})}}}

\def\twotwo(#1,#2,#3,#4){%
	\normalbaselines\m@th\offinterlineskip
	\vtop{\hbox{\htwobox({#1},{#2})}
	      \vskip-0.4pt
	      \hbox{\htwobox({#3},{#4})}}}

\def\Twotwo(#1,#2,#3,#4){%
	\normalbaselines\m@th\offinterlineskip
	\vtop{\hbox{\Htwobox({#1},{#2})}
	      \vskip-0.4pt
	      \hbox{\Htwobox({#3},{#4})}}}

\def\twooneone(#1,#2,#3,#4){%
	\normalbaselines\m@th\offinterlineskip
	\vtop{\hbox{\htwobox({#1},{#2})}
	      \vskip-0.4pt
	      \hbox{\fsquare(0.3cm,#3)}
	      \vskip-0.4pt
	      \hbox{\fsquare(0.3cm,#4)}}}

\def\Twooneone(#1,#2,#3,#4){%
	\normalbaselines\m@th\offinterlineskip
	\vtop{\hbox{\Htwobox({#1},{#2})}
	      \vskip-0.4pt
	      \hbox{\Fsquare(0.4cm,#3)}
	      \vskip-0.4pt
	      \hbox{\Fsquare(0.4cm,#4)}}}

\def\Twotwoone(#1,#2,#3,#4,#5){%
	\normalbaselines\m@th\offinterlineskip
	\vtop{\hbox{\Htwobox({#1},{#2})}
	      \vskip-0.4pt
	      \hbox{\Htwobox({#3},{#4})}
              \vskip-0.4pt
	      \hbox{\Fsquare(0.4cm,#5)}}}

\def\twotwoone(#1,#2,#3,#4,#5){%
	\normalbaselines\m@th\offinterlineskip
	\vtop{\hbox{\htwobox({#1},{#2})}
	      \vskip-0.4pt
	      \hbox{\htwobox({#3},{#4})}
              \vskip-0.4pt
	      \hbox{\fsquare(0.3cm,#5)}}}

\def\Threeoneone(#1,#2,#3,#4,#5){%
	\normalbaselines\m@th\offinterlineskip
	\vtop{\hbox{\Hthreebox({#1},{#2},{#3})}
	      \vskip-0.4pt
	      \hbox{\Fsquare(0.4cm,#4)}
              \vskip-0.4pt
	      \hbox{\Fsquare(0.4cm,#5)}}}

\def\threeoneone(#1,#2,#3,#4,#5){%
	\normalbaselines\m@th\offinterlineskip
	\vtop{\hbox{\hthreebox({#1},{#2},{#3})}
	      \vskip-0.4pt
	      \hbox{\fsquare(0.3cm,#4)}
              \vskip-0.4pt
	      \hbox{\fsquare(0.3cm,#5)}}}

\def\a{\fsquare(0.3cm){1}\addsquare(0.3cm)(2)\addsquare(0.3cm)(3)}

\def\b{\hbox{%
	\normalbaselines\m@th\offinterlineskip
	\vtop{\hbox{\fsquare(0.3cm){2}}\vskip-0.4pt\hbox{\fsquare(0.3cm){2}}}}}

\def\c{\hbox{\normalbaselines\m@th\offinterlineskip%
	\vtop{\hbox{\a}\vskip-0.4pt\hbox{\b}}}}


\dimen1=0.4cm\advance\dimen1 by -0.8pt
\def\ffsquare#1{%
	\fsquare(0.4cm,\hbox{#1})}

\def\naga{%
	\hbox{$\vcenter to 0.4cm{\normalbaselines\m@th
	\hrule\vfil\hbox to 1.2cm{\hfill$\cdots$\hfill}\vfil\hrule}$}}

\def\vnaga{\normalbaselines\m@th\baselineskip0pt\offinterlineskip%
	\vrule\vbox to 1.2cm{\vskip7pt\hbox to
\dimen1{$\hfil\vdots\hfil$}\vfil}\vrule}

\def\dvbox{\hbox{\normalbaselines\m@th\baselineskip0pt\offinterlineskip\vbox{%
	  \hbox{$\ffsquare 1$}\vskip-0.4pt\hbox{$\vnaga$}\vskip-0.4pt\hbox{$\ffsquare
N$}}}}

%
\vskip 0.5cm

In this note we sum up some results dealing with a level $r$ fusion algebra
${\cal F}_r({\gt g})$ for the semisimple Lie algebra ${\gt g}$. In fact we
will try to describe in details the fusion algebra for $\gt g ={\gt g}l(n)$ and
even for ${\gt g}=sl(2)$ leaving the general case as a collection of
likelihood conjectures. Our leading idea is to model the Kostant results
[Ko1], [Ko2] about the structure of the symmetric algebra of adjoint
representation of the Lie algebra ${\gt g}$ for the case of quantum deformation
of universal enveloping algebra $U_q({\gt g})$ at root of unity: $q^r=1$.

\bigbreak

{\bf \S1.~Multiplication in the $sl(2)$ fusion algebra. The Bethe ansatz
approach.}
\medbreak

We start with a consideration of the Lie algebra ${\gt g}=sl(2)$.
In this case a level $r$ fusion algebra ${\cal F}_r$ is defined as a finite
dimensional algebra over real numbers ${\bf R}$  with generators $\{v_j~|~
j=0,~1/2,~1,~3/2,~\ldots ,~r-2/2\}$ and the following multiplication rules
(the level $r$ Clebsch-Gordan series):
$$v_{j_1}{\widehat\otimes}v_{j_2}=\sum_{j=|j_1-j_2|,~j-j_1-j_2\in{\bf Z}}^
{\min (j_1+j_2,~r-2-j_1-j_2)}v_j. \eqno (1.1)
$$
Let us remark that the fusion rules (1.1) correspond to decomposing the tensor
product $V_{j_1}{\widehat\otimes}V_{j_2}$ of restricted representations [Ro]
$V_{j_1}$ and $V_{j_2}$ of the Hopf algebra $U_q(sl(2))$, when $q=\exp ({2\pi i
\over r})$, into the irreducible parts (see e.g. [Lu]). It is also well-known
(see e.g. [Ka])that the fusion algebra ${\cal F}({\gt g})$ is a commutative
and associative one.
\medbreak

{\bf Theorem 1.1.} Let us consider a decomposition of a product
$v_{j_1}{\widehat\otimes}\ldots {\widehat\otimes}v_{j_l}$
in the fusion algebra ${\cal F}_r$:
$$v_{j_1}{\widehat\otimes}v_{j_2}{\widehat\otimes}\ldots {\widehat\otimes}
v_{j_l}=\sum_ka_j(k)v_k.
$$
Then
$$a_j(k)=\sum_{\{\nu \}}\prod_{n\ge 1}\left(\matrix{P_{n,r}(\nu ,j)+m_n(\nu )
\cr m_n(\nu )\cr}\right),\eqno (1.2)
$$
where a summation in (1.2) is taken over all partitions $\nu =(\nu_1\ge\nu_2\ge
\cdots\ge 0)$ of the number $\sum_{s=1}^lj_s-k$ such that for all $n\ge 1$
the following inequalities are valid
$$P_{n,r}(\nu ,j):=\sum_{s=1}^l\min ~(n,2j_s)-2Q_n(\nu )-\max ~(2k+n-r+2,0)
\ge 0. \eqno (1.3)
$$
Here
$$\eqalignno{
&Q_n(\nu ):=\sum_{i\ge 1}\min ~(n,\nu_i)=\sum_{i\le n}\nu _i',\cr
&m_n(\nu ):=\nu _n'-\nu_{n+1}'}
$$
be the number of parts of the partition $\nu$ which are equal to $n$ and
$$\left(\matrix {m+n\cr n\cr}\right)={(m+n)!\over m!n!}
$$
be a binomial coefficient.

A proof of Theorem 1.1 is based on an investigation of the Bethe equations for
RSOS models [BR]. Note that validity of the equalities (1.2) for fixed $j_1,
\ldots j_l$ and for all $0\le k\le r-2$, are equivalent to a combinatorial
completeness (e.g. [Ki]) of the Bethe ansatz for RSOS models (for the case
${\gt g}=sl(2)$).

\qed

Now let us consider a simple example: $r=5,~j_1=\ldots =j_l=1$. We have two
``even'' representations in the fusion algebra ${\cal F}_5$, namely,
$V_0$ and $V_1$ and $V_1{\widehat\otimes}V_1=V_0+V_1$. Let us put
$V_1^{{\widehat\otimes}l}=a_lV_0+b_lV_1$. Then it is easy to see that
$a_{l+1}=b_l$
and $b_{l+1}=a_l+b_l=b_l+b_{l-1}$. Consequently, we have $a_l=F_{l-1}$ and
$b_l=F_l$, where $F_l$ be the $l$-th Fibonacci number. Hence from Theorem 1.1
we obtain
\medbreak

{\bf Corollary 1.2.} Let $F_l$ be the $l$-th Fibonacci number. Then we have
$$ 1)~~~~~~F_{l-1}=\sum_{\{\nu\}}\prod_{n\ge 1}\left(\matrix{P_{n,5}(\nu,
\bullet )+m_n(\nu )\cr m_n(\nu )\cr}\right), \eqno (1.4)
$$
where a summation in (1.4) is taken over all partitions $\nu $ such that
$$\eqalignno{
&i)~~~|\nu |=l\cr
&ii)~~P_{n,5}(\nu,\bullet ):=l\min ~(n,2)-2Q_n(\nu )-\max ~(n-3,0)\ge 0.}
$$
$$ 2)~~~~~~F_{l}=\sum_{\{\nu\}}\prod_{n\ge 1}\left(\matrix{{\widetilde P}
_{n,5}(\nu ,\bullet )+m_n(\nu )\cr m_n(\nu )\cr}\right), \eqno (1.5)
$$
where a summation in (1.5) is taken over all partitions $\nu $ such that
$$\eqalignno{
&i)~~~|\nu |=l-1 \cr
&ii)~~{\widetilde P}_{n,5}(\nu,\bullet ):=l\min ~(n,2)-2Q_n(\nu )-\max ~
(n-1,0)\ge 0.}
$$

Let us underline that formulae (1.4) and (1.5) give the different expressions
for the Fibonacci numbers. For example, a formula (4) gives for $F_7$ the
following expression in terms of riggid configurations
$$\matrix{\hbox{
          \normalbaselines\m@th\offinterlineskip
          \vtop{\hbox{{\Hthreebox( , , )}~0}
          \vskip-0.4pt
          \hbox{\Hthreebox( , ,)}
          \vskip-0.4pt
          \hbox{{\Htwobox( , )~4}}}}
&&\hbox{
          \normalbaselines\m@th\offinterlineskip
          \vtop{\hbox{{\Hthreebox( , , )}~0}
          \vskip-0.4pt
          \hbox{\Hthreebox( , , )}
          \vskip-0.4pt
          \hbox{{\Vtwobox( , )}~0}}}
&&\hbox{
          \normalbaselines\m@th\offinterlineskip
          \vtop{\hbox{{\Hthreebox( , , )}~0}
          \vskip-0.4pt
          \hbox{{\Twotwo( , , , )}~2}
          \vskip-0.4pt
          \hbox{{\Fsquare(0.4cm, )}~0}}}
&&\hbox{
          \normalbaselines\m@th\offinterlineskip
          \vtop{\hbox{{\Twotwo( , , , )}~0}
          \vskip-0.4pt
          \hbox{\Twotwo( , , , )}}}\cr
&  \cr
5&+&1&+&6&+&1&=&13,}
$$
where as the formula (1.5) gives the following one
$$\matrix{\hbox{
          \normalbaselines\m@th\offinterlineskip
          \vtop{\hbox{{\Hthreebox( , , )}~0}
          \vskip-0.4pt
          \hbox{{\Hthreebox( , , )}}}}
&&\hbox{
          \normalbaselines\m@th\offinterlineskip
          \vtop{\hbox{{\Hthreebox( , , )}~0}
          \vskip-0.4pt
          \hbox{{\Htwobox( , )}~3}
          \vskip-0.4pt
          \hbox{{\Fsquare(0.4cm, )}~1}}}
&&\hbox{
          \normalbaselines\m@th\offinterlineskip
          \vtop{\hbox{{\Twotwo( , , , )}~1}
          \vskip-0.4pt
          \hbox{\Htwobox( , )}}}\cr
& \cr
1&+&8&+&4&=&13.}
$$


It is possible to give the bijective proofs for identities (1.4) and (1.5).
Furthermore, using (1.4) and (1.5) one can construct two Fibonacci lattice
(e.g. [St1], [St2]). However, we do not assume to give here any combinatorial
details about very interesting combinatorial objects related with fusion
algebra: restricted Young tableaux, restricted Kostka-Foulkes polynomials,
restricted Littlewood-Richardson rule. All these things deserve a separate
publication.

Now let us return back to our example concerning with the fusion algebra
${\cal F}_5(sl(2))$. It seems a very interesting task to find the natural
$q$-analogs for identities (1.4) and (1.5).
We leave for another publication an exact
construction of such $q$-analogs, but here let us consider the following
well-known $q$-analog of the Fibonacci numbers $F_l(q)$. Namely, let us define
$$F_0(q)=0,~~F_1(q)=1,~~F_{l+1}(q)=qF_l(q)+F_{l-1}(q),~~l\ge 1.\eqno (1.6)
$$
It is easy to see that
$$\eqalignno{
&\sum_{l\ge 1}F_l(q)\cdot t^l={t\over 1-qt-t^2},~~{\rm and}\cr
&F_l(q)={\rm det}~|~\delta_{i,j}+\delta_{i,j-1}-q\delta_{i,j+1}|_{~1\le i,j\le
l-1}.}
$$
In order to understand better the relations between the $q$-Fibonacci numbers
(1.6) and fusion algebra ${\cal F}_5$, let us introduce an algebra
${\widetilde{\cal F}}_5=
\{v_0,v_1~|~v_1\cdot v_1=v_0+qv_1\} $.

Then it is easy to see that
$$v_1^l=F_{l-1}(q)\cdot v_0+F_l(q)\cdot v_1.\eqno (1.7)
$$
It is  possible to rewrite (1.7) as follows
$$\pmatrix{F_l\cr F_{l+1}\cr}=\pmatrix{0&1\cr 1&q\cr}\pmatrix{F_{l-1}\cr
F_l\cr}.
$$
Note that the matrices $\displaystyle{\pmatrix{0&1\cr 1&x\cr}}$ and
$\displaystyle{\pmatrix{0&1\cr 1&y\cr}}$ do not commute (if $x\ne y$):
$$\pmatrix{0&1\cr 1&x\cr}\pmatrix{0&1\cr 1&y\cr}=\pmatrix{1&y\cr x&1+xy\cr}
\ne \pmatrix{0&1\cr 1&y\cr}\pmatrix{0&1\cr 1&x}.
$$
However, it is easy to check that if we put $h(x)=\displaystyle{\pmatrix{1&x\cr
x&1+x\cr}}$ then $h(x)$ and $h(y)$ are commute. Now let us define a vacuum
vector $|0>:=\pmatrix{1\cr 0}$ and consider the following vectors
$$\eqalignno{
&h_n(x)|0>:=h(x_1)h(x_2)\ldots h(x_n)|0>=\pmatrix{a_n(x)\cr b_n(x)},&(1.8)\cr
&s_n(x)|0>:=2^nh_n\left({x\over 2}\right) |0>=\pmatrix{c_n(x)\cr d_n(x)}.}
$$
{}From what it was said above one can deduce that $a_n(x)$ and $b_n(x)$ are to
be
the symmetric functions on $x=(x_1,\ldots ,x_n)$. It is a simple exercise to
decompose these symmetric functions into a linear combination of the Schur
functions $s_{\lambda }(x)$.
\medbreak

{\bf Exercise 1.} Let us check:
$$\eqalignno{
&a_n(x)=1+\sum_{k=1}^{n-1}F_k\cdot s_{(1^{k+1})}(x),\cr
&b_n(x)=\sum_{k=1}^nF_k\cdot s_{(1^k)}(x),\cr
&c_n(x)=2^n+\sum_{k=1}^{n-1}2^{n-1-k}F_ks_{(1^{k+1})}(x),& (1.9)\cr
&d_n(x)=\sum_{k=1}^n2^{n-k}F_ks_{(1^k)}(x).}
$$
 Now let us calculate the values of $a_n(1)$ and $b_n(1)$. For this goal
we consider the following elements in the fusion algebra ${\cal F}_5(sl(2))$
$$\eqalignno{
&H(x)=v_0+xv_1=h(x)|0>,~~H:=H(1),\cr
&S(x)=2v_0+xv_1=\pmatrix{2&x\cr x&2+x}|0>,~~S:=S(1).}
$$
We must compute the coefficients of decompositions
$$\eqalignno{
&H^n:=a_nv_0+b_nv_1,\cr
&S^n:=c_nv_0+d_nv_1.}
$$
Here we give an answer only for $a_n$ and $c_n$.

{}From a definition it is clear that
$$a_{n+1}=a_n+b_n,~~b_{n+1}=a_n+2b_n,~~c_{n+1}=2c_n+d_n,~~d_{n+1}=c_n+3d_n.
$$
Consequently, if we put $\varphi (t)=\sum_{n\ge 1}a_nt^n$ and $\psi (t)=\sum
_{n\ge 1}c_nt^n$ then
$$\varphi (t)={t-t^2\over 1-3t+t^2},~~\psi (t)={2t-5t^2\over 1-5t+5t^2}.
\eqno (1.10)
$$
Using the generating functions (1.10), one can easily find ($n>0,~~k=3$):
$$\eqalignno{
&a_n=a_n(1)={2\over k+2}\sum_{m=0}^k\sin^2{(m+1)\pi\over k+2}\left( 2
\cos{(m+1)\pi\over k+2}\right)^{-2(n-1)},\cr
&c_n=c_n(1)={1\over 2}\sum_{m=0}^k\left( {\sqrt{4\over k+2}}\sin{(m+1)\pi\over
k+2}\right)^{-2(n-1)}.\cr
}
$$
In particular, we have the following equalities
$$\eqalignno{
&c_n=2^n+\sum_{k=1}^{n-1}2^{n-1-k}\pmatrix{n\cr k+1}F_k,\cr
&a_n=1+\sum_{k=1}^{n-1}\pmatrix{n\cr k+1}F_k,\cr
&d_n=\sum_{k=1}^N2^{n-k}\pmatrix{n\cr k}F_k.}
$$
Let us sum up the results of our computations in the fusion algebra
${\cal F}_5$. First of all, it was shown (for $r=5$) that a multiplicity of
a representation $V_0$
in the $n$-fold restricted tensor product
$S^{{\widehat\otimes }n}$ of the level $r$ symmetric algebra $S(=2V_0+V_1)$
is equal to the Verlinde number $V(r-2,n)$:
$$[V_0:S^{{\widehat\otimes}n}]=V(r-2,n):={1\over 2}\sum_{m=0}^{r-2}(S_{0m})
^{-2(n-1)},
$$
where
$$S_{jm}={\sqrt{4\over r}}~\sin{(j+1)(m+1)\pi\over r}.
$$
After this we proved an existence of a ${\gt g}l(n)$-module
${\cal V}^{(0)}$
such that ${\rm dim}~{\cal V}^{(0)}=V~(r-2,n)$ and computed it character:
$${\rm ch}{\cal V}^{(0)}=\sum_{\lambda }a_{\lambda }W_{\lambda },
$$
where $a_{\lambda }\in{\bf Z}_+,~~l(\lambda )\le n,~~l(\lambda ')\le
{\displaystyle{r-3\over2}}$ and $W_{\lambda }$ be an irreducible
representation of ${\gt g}l(n)$ with the highest weight $\lambda $.
\bigbreak

{\bf \S 2.~Verlinde character.}
\medbreak

In this section we study a decomposition of a product of some distinguish
elements in the fusion algebra ${\cal F}_r:={\cal F}_r(sl(2))$. Thus we start
with a definition of these elements. Let us define a level $r$ symmetric
algebra ${\cal S}_r$ and a module of level $r$ harmonic polynomials
${\cal H}_r$ as follow
$$\eqalignno{
&{\cal S}_r=\sum_{k=0}^{r-3\over 2}\left({r-1\over 2}-k\right) x^k\cdot V_k\in
{\cal F}_r[x],\cr
&{\cal H}_r=\sum_{k=0}^{r-3\over 2}x^k\cdot V_k\in{\cal F}_r[x].&(2.1)}
$$
In sequel we will assume that $r\equiv 1({\rm mod}~2)$. Our nearest aim is to
find the coefficients in the following decompositions
$$\eqalignno{
&{\cal S}_r^{{\widehat\otimes}n}=\sum_{k=0}^{r-3\over 2}a_{k,n}(x)\cdot V_k,
&(2.2)\cr
&{\cal H}_r^{{\widehat\otimes}n}=\sum_{k=0}^{r-3\over 2}b_{k,n}(x)\cdot V_k.}
$$
But at the beginning we find the values $a_{k,n}(1)$ and $b_{k,n}(1)$.
\medbreak

{\bf Theorem 2.1.}~~We have
$$\eqalignno{
&a_{0,n}(1)={1\over 2}\sum_{m=0}^{r-2}\left({\sqrt{4\over r}}~\sin{(m+1)
\pi\over r}\right)^{2-2n},&(2.3)\cr
&b_{0,n}(1)={2\over r}\sum_{m=0}^{r-2}\sin^2{(m+1)\pi\over r}\left( 2
\cos{(m+1)\pi\over r}\right)^{2-2n}.}
$$
Sketch of a proof. We start with a solution of a local problem, namely, we
want to find a connection matrix $M(x)$ (resp. $N(x)$) such that
$$\eqalignno{
&{\cal S}_r^{{\widehat\otimes}(n+1)}=M(x)\cdot {\cal S}_r^{{\widehat\otimes}n},
&(2.4)\cr
&{\cal H}_r^{{\widehat\otimes}(n+1)}=N(x)\cdot{\cal H}_r^{{\widehat\otimes}n}.}
$$
\medbreak

{\bf Proposition 2.2.}~~The connection matrices $M(x)=(m_{ij}(x))$ and
$N(x)=(n_{i,j}(x))$ have the following matrix elements
($1\le i,j\le{r-1\over 2}$)
$$\eqalignno{
&m_{i,j}(x)=x^{i-j}\left\{\sum_{k=0}^{\min (2i-2,~r-2j)}\left({r-1\over 2}+
i-j-k\right) x^k\right\} ,&(2.5)\cr
&{\rm if}~~i\ge j~~{\rm and}~~m_{ij}=m_{ji},~~{\rm if}~~i\ge j;\cr
&n_{ij}(x)=x^{i-j}\left\{\sum_{k=0}^{\min (2i-2,~r-2j)}x^k\right\} ,&(2.6)\cr
&{\rm if}~~i\ge j~~{\rm and}~~n_{ij}=n_{ji},~~{\rm if}~~i\ge j.}
$$
The statement of Proposition 2.2 may be verified by direct computation using
the definitions (2.1).

\qed

\medbreak

{\bf Corollary 2.3.}~~We have
$$m_{ij}(1)={1\over 2}(2i-1)(r+1-2j),~~{\rm if}~~1\le i\le j\le{r-1\over 2},
$$
and $m_{ij}(1)=m_{ji}(1)$ if $i\ge j$.

The next step of a proof of Theorem 2.1 is the following observation.
\medbreak

{\bf Proposition 2.4.}~~Matrix $M(1)$ admits a decomposition
$$M(1)=T\cdot C^{-1},
$$
where $C=(c_{ij}),~~c_{ij}=2\delta_{i,j}-\delta_{i,j+1}-\delta_{i,j-1}~~
(1\le i,j\le{r-1\over 2})$ be the Cartan matrix of tipe $A_{r-1\over 2}$ and
$T=(t_{ij})$ be a lower triangle matrix with elements
$$\eqalignno{
&t_{11}={r+1\over 2}={\rm det}~C,\cr
&t_{ij}=r~~{\rm and}~~t_{1j}=-\left({r+1\over 2}-j\right) ,~~{\rm if}~~
2\le j\le{r-1\over 2},\cr
&t_{ij}=0~~{\rm otherwise}.}
$$
Using Proposition 2.4 we may find the spectra of matrix $M(1)$ and solve a
recurrence relation (2.4) for ${\cal S}_r^{{\widehat\otimes}n}$.
As a result, we obtain the first formula of (2.3).

\qed

Now we want to construct the representations ${\cal V}^{(k)}$ and
${\cal W}^{(k)}$,
$1\le k\le{r-1\over 2}$, of the Lie algebra ${\gt g}l(n)$ such that
$${\rm dim}~{\cal V}^{(k)}=a_{k,n}(1),~~{\rm dim}~{\cal W}^{(k)}=b_{k,n}(1).
$$
For this aim we use the following observation
\medbreak

{\bf Proposition 2.5.}~~If the matrices $M(x)$ and $N(x)$ are defined by the
formulae (2.5) and (2.6), then they satisfy the commutation relations
$$\eqalignno{
&M(x)\cdot M(y)=M(y)\cdot M(x),\cr
&N(x)\cdot N(y)=N(y)\cdot N(x),}
$$
Finally, let us define a vacuum vector
$|0>=(1,0,\ldots ,0)^t\in R^{r-1\over 2}$
and consider the following vectors
$$\eqalignno{
&M(x_1)\ldots M(x_n)~|0>=(\chi_{_{1,n}}(x),\ldots ,\chi_{_{{r-1\over 2},n}}
(x))^t,\cr
&N(x_1)\ldots N(x_n)~|0>=(\varphi_{_{1,n}}(x),\ldots ,\varphi_{_{{r-1\over
2},n}}(x))^t.}
$$
It is clear from our construction that
$$\eqalignno{
&\chi_{_{k,n}}(1)=a_{k,n}(1)~~{\rm and}~~\varphi_{_{k,n}}(1)=b_{k,n}(1),\cr
&\chi_{_{k,n}}(x,\ldots ,x)=a_{k,n}(x)~~{\rm and}~~\varphi_{_{k,n}}(x,\ldots
,x)
=b_{k,n}(x).}
$$
\medbreak

{\bf Theorem 2.6.}~~The symmetric functions $\chi_{_{k,n}}(x)$ and
$\varphi_{_{k,n}}(x),~~1\le k\le{r-1\over 2}$, may be expressed as the linear
combinations of the Schur functions with positive integer coefficients, i.e.
$$\chi_{_{k,n}}(x)=\sum_{\lambda }a_{\lambda }s_{\lambda }(x),
$$
where for all partitions $\lambda$ we have $a_{\lambda }\in{\bf Z}_+$ and if
$a_{\lambda }\ne 0$ then $l(\lambda )\le n$ and $l(\lambda ')\le{r-3\over 2}$.

\qed

It seems an interesting task to define a natural action of the Lie group
$GL(n)$ on the space $H^0({\cal M}_n,L^{\otimes (r-2)})$ such that
$${\rm char}~H^0({\cal M}_n,L^{\otimes (r-2)})=\chi_{_{0,n}}(x).
\eqno (2.7)
$$
Probably, such action may be extracted from [Kh].
\medbreak

{\bf Acknowledgements.}~~Most of this work was done during my visit to the
Isaac Newton Institute for Mathematical Sciences. I thank B.Feigin, P.Mathieu,
J.Weitsman and E.Corrigan for very helpfull discussions.
\bigbreak

{\bf References.}
\medbreak
\item{[Bo]} R.Bott, Int. Journ. Mod. Phys. A, 6, (1991), 2847.
\item{[Br]} V.V.Bazhanov, N.Yu.Reshetikhin, Int. Journ. Mod. Phys. A, 4,
(1989), 115.
\item{[FZ]} V.Fateev, A.Zamolodchikov, Soviet Journ. Nucl. Phys., 43, (1986),
630.
\item{[JW]} L.C.Jeffrey, J.Weitsman, Topic structure on the moduli space of
flat connections on a Riemann surface : Volumes and the moment map, prepr.
IASSNS-HEP-92/25.
\item{[Ka]} V.G.Kac, Infinite dimensional Lie algebras, Cambridge University
Press, 1990.
\item{[Ki1]} A.N.Kirillov, Journ. Geom. and Phys., 5, (1988), 365.
\item{[Ki2]} A.N.Kirillov, Journal of Soviet Math., 36, (1987), 115.
\item{[Ki3]} A.N.Kirillov, Dilogarithm identities, partitions and spectra in
conformal field theory, in preparation.
\item{[Kh]} T.Kohno, Topology, 31, (1992), 203.
\item{[Ko1]} B.Kostant, Amer. Journ. Math., 81, (1959), 973.
\item{[Ko2]} B.Kostant, Amer. Journ. Math., 85, (1963), 327.
\item{[Lu]} G.Lusztig, Journ. Amer. Math. Soc., 3, (1990), 257.
\item{[Ro]} M.Rosso, Commun. Math. Phys., 124, (1989), 307.
\item{[St1]} R.Stanley, Fibonacci Q., 13, (1975), 215.
\item{[St2]} R.Stanley, Europ. J. Combinatorics, 11, (1990), 181.
\item{[Sz]} A.Szenes, Int. Math. Research Notices, 7, (1991), 93.
\item{[Th]} M.Thaddeus, Journ. Diff. Geometry, 35, (1992), 131.
\item{[Ve]} E.Verlinde, Nucl. Phys., B300, (1988), 351.

%
\end